\documentclass[aps,prl,twocolumn,groupedaddress,showpacs,floatfix,letterpaper]{revtex4}

\usepackage{amsfonts}
\usepackage{amssymb}
\usepackage{amsbsy}
\usepackage{amsmath}
\usepackage{amstext}
\usepackage[mathscr]{eucal}

\usepackage{graphicx,color}

\usepackage{verbatim}

\begin{document}

\title{Instabilities and ``phonons" of optical lattices in hollow optical fibers}

\author{N.~K.~Wilkin and J.~M.~F.~Gunn}

\affiliation{School of Physics and Astronomy, University of
Birmingham, Edgbaston, Birmingham. B15 2TT. U.~K.}

\begin{abstract}
Instabilities are predicted for a sufficiently long hollow photonic optical fiber, or ``cavity", containing a one dimensional Bose-gas in the presence of a classical, far red-detuned, confined weak electromagnetic mode. We examine both a single beam with Bose gas (a type of Brillouin instability) and the case of a standing wave, or optical lattice.  The instabilities of these driven systems have pronounced spatial structure, of combined modulational instabilities in the electromagnetic and Bose density fields. Near the critical wave vectors for the instability the coupled modes of the BEC and light can be interpreted as ``phonons" of the optical lattice. We conjecture these spatially-structured instabilities for the optical lattice, which we predict at weak fields, develop into the source of spatially homogeneous heating predicted for strong fields.
\end{abstract}
 \maketitle

Systems where optical fields do not just determine the potential energy which cold atoms experience (via the quasi-static averages of the fields), but actually are dynamical participants, have recently become prominent. These range from symmetry breaking transitions\cite{Ess}, strongly interacting photons\cite{Dem} and collective dynamics\cite{Ritsch}.  

There is an even simpler context in which light may not be a bystander. Given an optical lattice is a driven system, it is natural to ask if it is stable or whether there are parameter values where the spatial periodic (and constant temporal) order disintegrates. In this Letter we address this issue in the simplest case: where the optical field is weak and both the atoms and photons are constrained to move in one-dimension, for example in hollow optical fibres. While the frequency scales of the optical and atomic density fields are very disparate - typically eleven orders of magnitude or more - this leaves open the existence of modulational instabilities of light on frequency scales comparable with collective modes in the Bose gas, and hence cooperative effects. 

The transportation of atoms along hollow optical fibres by light was suggested\cite{Ol,Zol}  twenty years ago, with the first experiments\cite{Cor} occurring not long after.  The invention\cite{Pr} of hollow photonic optical fibres gave new impetus to this field due to the much lower attenuation of light (attenuation being $\gtrsim1.5$dB/Km\cite{Rusatt}, with commercial values being $\sim100$dB/Km). Demonstration\cite{Pr2} soon followed of propulsion of micron-sized particles along them using the optical modes in the fibres. Recently\cite{Pr3} these methods have developed to be able to transport such particles over distances of several metres.  Atoms\cite{Tak}  have been guided along photonic fibres, with  motivation ranging from exploiting the protected environment of the fibre for the atoms, to potential applications such as nonlinear optics at ultralow intensities\cite{Gae}. Cold atoms have been loaded into photonic fibres with increasing success\cite{Ket,Bac}. Mysteriously, the transportation of BECs (as against single atoms) in fibres has proved so far to be a challenge to achieve without significant atomic attenuation, especially in a continuous, as against a pulsed\cite{Kai}, mode.

We will be working in the regime of wave vectors where the Bogoliubov approximation is a good description of the excitation spectrum of the one dimensional Bose gas, and the Thomas-Fermi approximation is appropriate.  We may then use  a hydrodynamic formulation derived from the Gross Pitaevskii (GP) equation.  The atom number density, $\rho(x,t)$, (which is dimensionally the three-dimensional density, including an inverse factor of the inverse cross-sectional area of the transverse wave function) and $S(x,t)$ which contains the velocity potential and chemical potential,  are related to the ``condensate" wave function, $\phi(x,t)$, by $\phi=\sqrt{\rho}{\rm e}^{{\rm i}S}$. The familiar Bernoulli-like and continuity equations in the presence of an electric field, with magnitude $E (x,t)$,  take the form respectively  ($A_x = \partial A/\partial x$ etc):
\begin{eqnarray}
 (S_x)^2  +{\tilde \rho}+ S_t -\textstyle{\frac{1}{2}}\alpha \langle E^2 \rangle &=& 0, \label{eq:Bern}\\
{\tilde \rho}_t + \partial_x ({\tilde \rho} S_x) &=& 0.\label{eq:Cty}\end{eqnarray}
We have used dimensionless length, energy and time variables formed from the healing length and the Hartree potential, $g{\overline \rho}$, with $g$ being the coupling constant in the GP equation.  ${\overline \rho}$ is the average number density of atoms;  ${\tilde \rho} = \rho/{\overline \rho}$ is the dimensionless atom number density with mean value unity. The last term in Eqn.(\ref{eq:Bern}) is the interaction with the optical field, where $\alpha = \alpha_0 /(g{\overline \rho})$, with $\alpha_0$ being the atomic polarisability.  $\langle  E^2 \rangle$ is the time-average, over high frequencies, of the optical field intensity.

The equation for the electromagnetic field, $E(x,t)$, using the same dimensionless variables,  is 
\begin{equation}
-\partial_x^2 E + {\overline {\rm c}}^{-2}\partial_t^2 \left(1+{\tilde \alpha} ({\tilde \rho} -1) \right)E=0,
\label{eq:emag}\end{equation}
where ${\tilde \alpha} = 4\pi \alpha_0 {\overline \rho}/(1+4\pi \alpha_0 {\overline \rho})$, ${\overline {\rm c}}= {\rm c}/[(1+4\pi \alpha{\overline \rho})^{1/2}c_{\rm s}]$ (accommodating the effect of the average medium formed by the atoms) with $\rm c$ is the speed of light in vacuo, and $c_{\rm s}$ is the speed of sound in a BEC corresponding to the average density, $\overline \rho$. 

Assuming the fiber can be populated with light and atoms, the simplest case is whether {\em one} light beam and uniform gas initially placed in the fiber are stable.  Equations (\ref{eq:Bern}), (\ref{eq:Cty}) and (\ref{eq:emag}) are satisfied by the initial configuration ${\tilde \rho} = 1$, $S(t) = -t(1+\frac{1}{2} \alpha' \langle  E^2 \rangle) $ (i.e. $-S/t=\mu$, the chemical potential) and $E(x,t) = \sqrt{2\langle  E^2 \rangle} \cos (Kx-\Omega t)$, where ${\overline {\rm c}}^2K^2 = \Omega^2$. Let us define small fluctuations around that solution by ${\tilde \rho} = 1 + n(x,t)$ and  $E (x,t) = E_0 (x,t) + \varepsilon(x,t)$, 
where $E_0 (x,t) = \cos (Kx-\Omega t)$, i.e. we have normalised the electric fields. Then the linearised equations are, defining $\alpha' = \alpha \langle  E^2 \rangle$:
\begin{eqnarray}
- n_{tt}+n_{xx} - \alpha' \partial_x^2 (E_0 \varepsilon) &=& 0,\\
- \varepsilon_{xx} + {\overline {\rm c}}^{-2} \varepsilon_{tt} +{\tilde \alpha}{\overline {\rm c}}^{-2}\partial_{t}^2 (n E_0) &=&0. \end{eqnarray}
The equations have the form of two coupled, and distributed, parametric oscillators. Note the parametric driving terms (from the external optical field) are in the coupling of the electromagnetic and density fields. These equations lead to a Brillouin instability, although the initial conditions, and hence treatment,  are unconventional. 

\begin{figure}[h]
\includegraphics[width=.9\linewidth]{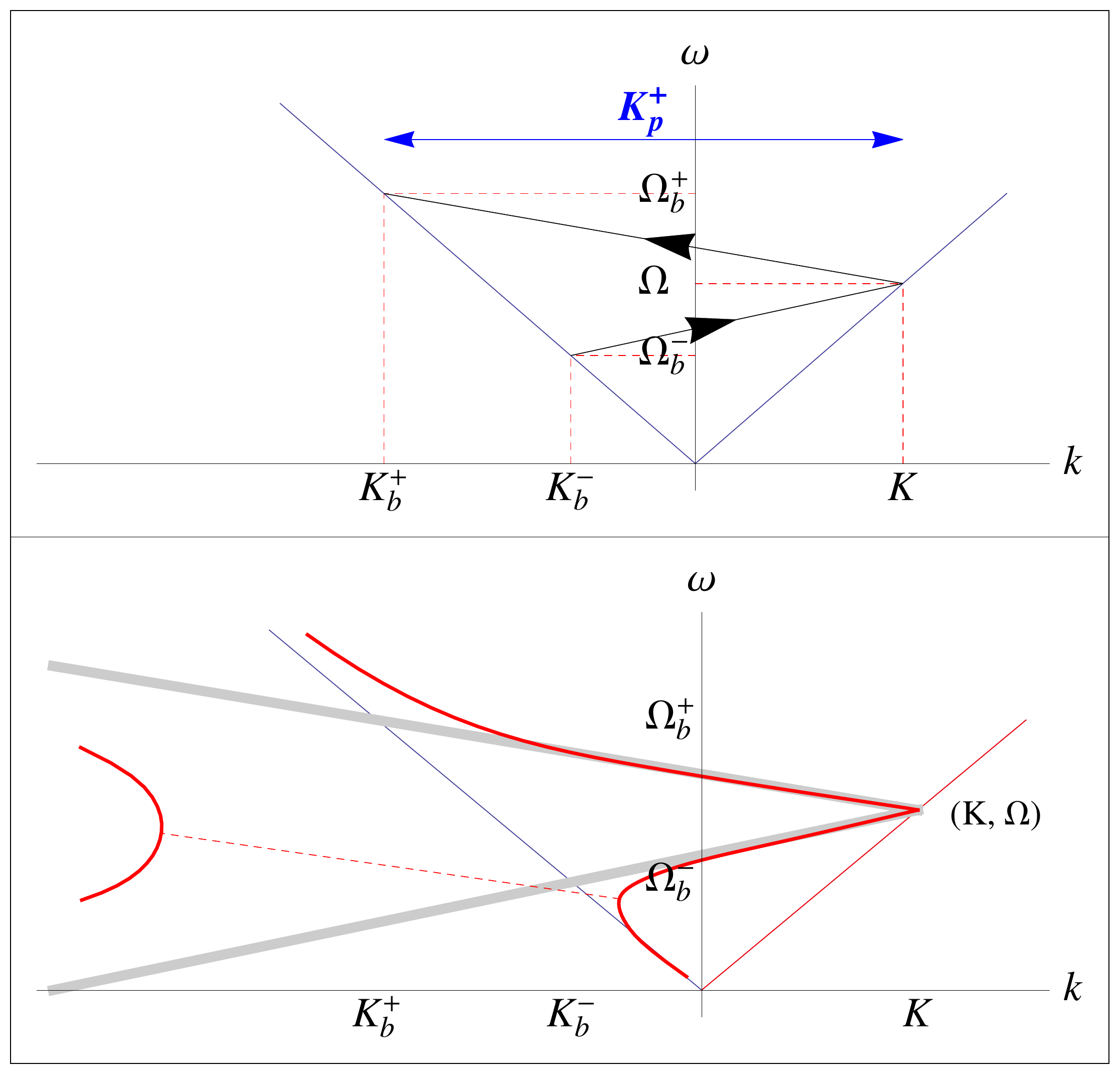}
\caption{The upper figure shows the two ``kinematic points", showing the initial photon (wave vector $K$) and the Brillouin-scattered photons, $K_{\rm b}^\pm$ and phonons, $K_{\rm p}^\pm$, not to scale. The lower figure is a schematic indicating the coupled modes (undamped modes indicated by the full red curves), with the real part of $\omega$ for the unstable mode indicated by the dashed line. Compare with the full calculation in Fig.(\ref{fig:brillouin}).}
\label{fig:kinp}\end{figure}

The nature of the instability, both spectrally and spatially, is best understood in $k$-space. The most pronounced effect of the coupling between the atom density field and the electromagnetic field is expected to be where there is a degeneracy between the energy and momentum of the initial and final states. The  ``kinematic points" (KPs) where this occurs are shown in Fig.(\ref{fig:kinp}). The unstable region in $k$-space is much larger than the difference of wave vector of the two modes (of order $1/{\overline c}$), so we henceforth talk of {\em one} KP with photons $K_{\rm b}=-K$ and the phonons having wave vectors of $K_{\rm p} = 2K$. 

\begin{figure}
\includegraphics[width=.9\linewidth]{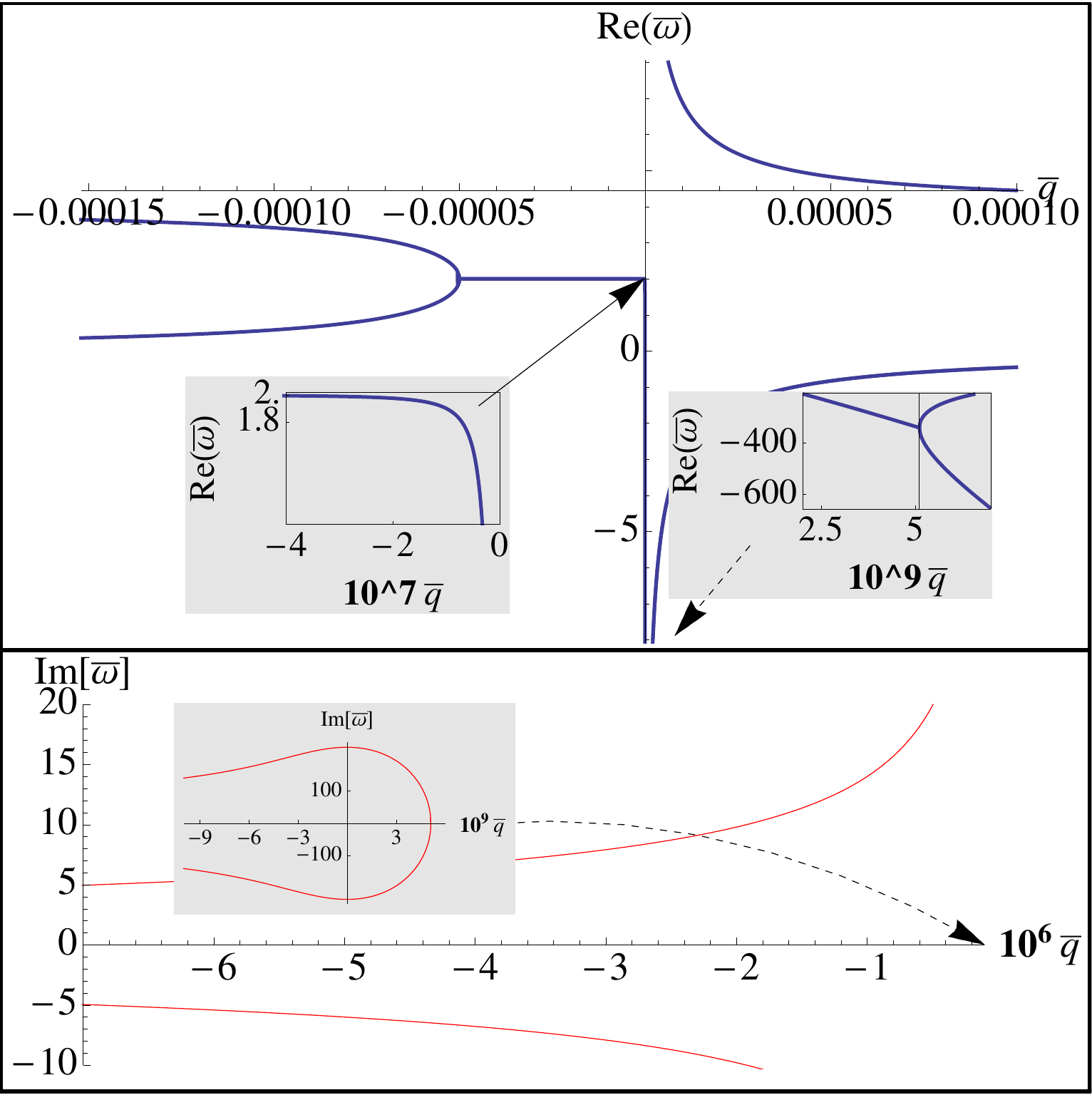}
\caption{The real and imaginary parts of the frequency for the modes with an initial plane electromagnetic wave. The insets show fine detail in $\overline q$, where the unstable branch is seen to be not quite parallel to the vertical (Re$\overline \omega$) axis.}\label{fig:brillouin}
\end{figure}

As the coupling parameters are small, $\alpha'\ll 1$ and ${\tilde \alpha}\ll 1$, any Brillouin-type instability will be in the vicinity of the kinematic point, where we represent $n$ and $\varepsilon$ by
\begin{eqnarray*}
n(x,t) &=& n{\rm e}^{{\rm i} [(2K-q)x-(2K -\omega)t]} + \hbox{c. c.},\\
\varepsilon (x,t) &=& \varepsilon_0{\rm e}^{{\rm i} [(-K +q)x- (\Omega -\{ K_{\rm p} -\omega\}) t]} + 
\hbox{c. c.},\end{eqnarray*}
where the deviation from the KP is denoted by $q$ and $\omega$.  We approximate $(\Omega-K_{\rm p})/{\overline c} \simeq K_{\rm b}$ and only include the dominant contribution in the coupling terms. Then, scaling $q = K {\overline q}$ and $\omega = K{\overline \omega}$, we see: 
\begin{align}[({\overline q}-{\overline \omega}) +\textstyle{\frac{1}{4}}({\overline \omega}^2 - {\overline q}^2)]n_0 +  \alpha' \varepsilon_0^* &= 0,\\
[ -2 ({\overline q}
+{\overline \omega}/{\overline {\rm c}} )
+({\overline q}^2 -
{\overline \omega}^2
/{\overline {\rm c}}^2
)
]
\varepsilon_0^*
-{\tilde \alpha} n_0 &= 0.\label{eq:epsi}
\end{align}
We solve these equations both numerically and asymptotically\cite{Simm} using the two small parameters $V_2={\tilde \alpha} \alpha '/4$, and ${\overline {\rm c}}\thinspace{}^{-1}$. Let us assume $\tilde \alpha \simeq 10^{-3}$, i.e. ${\overline \rho} \simeq 5 \times10^{14}{\rm cm}^{-3}$ and $\alpha' \simeq 0.1$ (to ensure weak optical lattice is a valid description), which would imply, for example, approximately a decrease of the laser power in Ref. [\onlinecite{Kai}] by a factor of 100 and an increase of the beam radius (and of the core of the fiber) by a factor of 10. Then we find that $V_2 \sim 10^{-5}$ and ${\overline {\rm c}}\thinspace{}^{-1} \sim 10^{-11}$. The resulting dispersion relation, ${\overline \omega} ({\overline q})$, from the numerical solution of these equations is shown in Fig.(\ref{fig:brillouin}).  

In the stable region, the distortion of the modes as mixtures of photon and phonon, but {\em not} a polariton) near the instability for ${\overline q}>0$ is visible, with the generic (for rank 2 non-Hermitian matrices) square root form, ${\overline \omega} ({\overline q}) \sim (q-q_{\rm c})^{1/2}$, which generates an imaginary part in the unstable region and a modification to the real part as the instability is approached. There is also an upper, stable, branch which crosses over from being phonon-like to photon-like.  The unstable region extends further in the negative $\overline q$ direction with the critical value of the scaled wave vector being ${\overline q}^-_{\rm c} \simeq -2V_2$ and ${\overline \omega}_{\rm c}\simeq 2$ with the two phonon branches separating after the instability. In the positive $\overline q$ direction, we find 
$${\overline q}^+_{\rm c} \simeq 3 (2V_2/{\overline c}^2 )^{1/3}\quad \hbox{and} \quad {\overline \omega}^+_{\rm c} \simeq - 2 (2V_2 {\overline c})^{1/3},$$
with one photon and one phonon branch emerging.  For the case of ${\overline q}$ being just within the instability region, we find: ${\overline q} \simeq  {\overline q}_{\rm c}^\pm (1-\delta) \quad {\overline \omega}\simeq {\overline \omega}_{\rm c}^\pm(1\pm {\rm i}\delta^{1/2})$. The main weight of the eigenfunction (there is only one at the instability, associated with the non-Hermiticity) is in the optical field near both instabilities, overwhelmingly for ${\overline q}^{\thinspace +}_{\rm c}$,   and to a lesser extent near ${\overline q}^{\thinspace -}_{\rm c}$:
$$n_0({\overline q}^{\thinspace +}_{\rm c})/\varepsilon^*_0({\overline q}^{\thinspace +}_{\rm c}) \simeq - \alpha'(2V_2{\overline c})^{-2/3}, n_0({\overline q}^{\thinspace -}_{\rm c})/\varepsilon^*_0({\overline q}^{\thinspace -}_{\rm c}) \simeq \alpha'.$$
The counter propagating scattered light and phonons are reminiscent to an optical lattice. If the initial state were an optical lattice, would this be stable?  We will now see, in the case of a ``weak" lattice, $n_0(x)\ll 1$,  that the change in the dispersion relation (due to the optical band gaps from the periodic $n_0(x)$) weakens, {\em but does not destroy} the instability. 

An unperturbed optical lattice emerges as a solution to Eqns. (\ref{eq:Bern}), (\ref{eq:Cty}) and (\ref{eq:emag}) where: the atomic density, ${\tilde \rho} = 1 + n_0(x)$, is time-independent and the spatial average of $n_0(x)$ is zero; there are no currents, so $S(t)$ is independent of $x$; and the electromagnetic field is a standing wave, $E(x,t) = {\cal E}(x) \cos (\Omega t)$, of fixed frequency $\Omega$. The continuity equation then vanishes identically. Specialising to the Thomas-Fermi limit, and defining the {\em spatial} average of $\langle  E^2 \rangle$ to be $\overline{\langle  E^2 \rangle}$, the equations for $n_0(x)$, $S(t)$ and $E(x,t)$  become
\begin{eqnarray} 
1+ n_0(x) + S_t - {\textstyle \frac{1}{2}} \alpha' (\overline{\langle  E^2 \rangle} + \langle  E^2 \rangle -   \overline{\langle  E^2 \rangle} )&=&0,\label{eq:Berna}\\
-\partial_x^2 E + {\overline {\rm c}}^{-2}\partial_t^2 \left[1+{\tilde \alpha} n_0(x)] \right)E &=&0.
\label{eq:emaga}
\end{eqnarray}
If we choose fixed average density, $\overline{{\tilde \rho}} =1$, (as against fixed chemical potential) then the solution to Eqn. (\ref{eq:Berna})  is
\begin{eqnarray} S(t)  &=& -t(1 - {\textstyle \frac{1}{2}} \alpha' \overline{\langle  E^2 \rangle} ),\nonumber \\ 
n_0(x) &=&  {\textstyle \frac{1}{2}}\alpha' (\langle  E^2 \rangle -   \overline{\langle  E^2 \rangle} ) 
=  {\textstyle \frac{1}{4}}\alpha' [{\cal E}^2(x) - \overline{{\cal E}^2}]\!\! , \label{eq:n0}\end{eqnarray}
using the definition of $\cal E$. Note that the chemical potential, $\mu = -S/t$, is adjusted to ensure the density retains the same value in the presence of the optical field. We may now substitute the expression for $n_0(x)$, and $E(x,t)={\cal E}(x) \cos (\Omega t)$, into Eqn. (\ref{eq:emaga}) to obtain
\begin{equation}
-{\cal E}_{zz} + {\cal E}+A {\cal E}^3  =0,
\label{eq:emagnl}\end{equation}
where $z= x\Omega (1- {\tilde \alpha}\alpha' \overline{{\cal E}^2}/2)/{\overline {\rm c}}$ and $A= {\tilde \alpha}\alpha' \overline{{\cal E}^2}/(2- {\tilde \alpha}\alpha' \overline{{\cal E}^2})$. Eqn. (\ref{eq:emagnl}) is the Duffing equation\cite{Duffing}, providing an exact expression for the wave vector, $K_{\rm lat}$, of the light in the optical lattice (assuming the frequency, $\Omega$, is fixed by the lasers), using the equivalence of the period of the Duffing oscillator and the wave length of the optical lattice:
$$K_{\rm lat} = K_0 (1+{\tilde \alpha} -\alpha'{\tilde \alpha}\overline{{\cal E}^2}/4)^{1/2}(1+A{\cal E}^2_{\rm m}/2)^{1/2}[K(k)]^{-1}, $$
where $K_0$ is the vacuum wave vector corresponding to $\Omega$ and $K(k)$ is the complete elliptic integral with $k^2 = - A{\cal E}^2_{\rm m}/(2 + A{\cal E}^2_{\rm m})$. We may deduce the corresponding solution to Eqn. (\ref{eq:emagnl}) as being:
\begin{eqnarray}
{\cal E}(x)\!\!\!\!\!\!\! &=&\!\!\!\! \!\!{\cal E}_{\rm m} \thinspace {\rm cn} [(1+{\textstyle \frac{1}{2}}A{\cal E}^2_{\rm m})^{1/2}(1+{\tilde \alpha}- {\textstyle \frac{1}{4}}\alpha'{\tilde \alpha}\overline{{\cal E}^2})^{1/2} K_0 x, k]\nonumber\\
&\simeq& \!\!\!\!\!\!{\cal E}_{\rm m} (1+{\textstyle \frac{5}{32}}\alpha'{\tilde \alpha}{\cal E}^2_{\rm m})\nonumber\\
&\phantom {\simeq {\cal E}_{\rm m}} &\cos [(1+{\textstyle \frac{1}{2}}{\tilde \alpha} + {\textstyle \frac{3}{8}}A{\cal E}^2_{\rm m} - {\textstyle \frac{1}{8}}\alpha'{\tilde \alpha}\overline{{\cal E}^2} )K_0 x],
\end{eqnarray}
where cn$(z,k)$ is a Jacobian elliptic function, we have approximated by taking only the first term in the two small parameters ${\tilde \alpha}$ and $\alpha'{\tilde \alpha}$ and using the first term of the Fourier series\cite{NISTFou} for cn$(z,k)$ (consistently with the previous approximation). To compare with  a solution of the form $E_{\rm lat}(x,t)= 2E_0\cos (Kx)\cos (\Omega t)$, using the definitions of $\overline{{\cal E}^2}$ and ${\cal E}_{\rm m}$, we see 
\begin{equation}E_{\rm lat}(x,t) = 2E_0 \cos [K_{\rm lat}^{\rm lin} x] \cos(\Omega t),\label{eq:optlatlin}\end{equation}
where we have approximated: $K_{\rm lat}\simeq K^{\rm lin}_{\rm lat} = (1+{\textstyle \frac{1}{2}}{\tilde \alpha}  + {\textstyle \frac{1}{8}}\alpha'{\tilde \alpha}E_0^2)K_0$. The interpretation of the second and third terms multiplying $K_0$, are:  the effect of the average refractive index of the atomic medium; and the coincidence of the maxima of $n_0(x)$ with the antinodes of $E(x,t)$, providing an enhanced effective refractive index, respectively.  The need for, and interpretation of, the self-consistency of the periodicities of atomic density and optical field was discussed theoretically and experimentally in the strong optical lattice limit in Refs. [\onlinecite{Bill1,Bill2}].

We will see there is a complex arrangement of instabilities which combine density waves and modulation of the optical lattice field. Near the instabilities, the stable combined modes may be thought of as ``phonons" of the optical lattice, with the optical field modulated at same frequency as the atomic density field. We now use Eqns. (\ref{eq:n0}) and $K_{\rm lat} $ as a basis to expose descendants of the Brillouin instability. 

Phonons propagating in {\em both} directions must be included with wave numbers, $k$, near $k=\pm 2K$.   We define the amplitudes of the left- and right-propagating photons and phonons as :
\begin{eqnarray}
\varepsilon^*  &=& {\rm e}^{{\rm i}(\Omega-\omega)t}( {\rm e}^{-{\rm i}(K-q)x}\varepsilon^*_+ +  {\rm e}^{-{\rm i}(-K-q)x}\varepsilon^*_-)\!\! ,\\
n  &=& {\rm e}^{-{\rm i}(2K -\omega) t} ({\rm e}^{{\rm i}(2K+q)x} n_+ + {\rm e}^{{\rm i}(-2K+q)x} n_- )\!\! .
\end{eqnarray}
The two phonon states, $2K+q$ and $-2K+q$, are coupled\cite{Usf} by Bragg reflection from a higher order contribution to $n_0(x)$, of order $\alpha^{\prime 2}$, which we may neglect in the following analysis. 

We utilise the instability being near the optical lattice wave vector as in the Brillouin instability above. It is convenient to scale $q= 2K {\overline q}$ and $\omega = 2K{\overline \omega}$, we may omit terms in ${\overline q}^2$ and use the expression for $K^{\rm lin}_{\rm lat}$. The resulting equations with coefficients of the Fourier components ${\rm e}^{{\rm i}(2K+q)}$ and ${\rm e}^{{\rm i}(-2K+q)}$ from the equation for $n(x,t)$ and ${\rm e}^{-{\rm i}(K-q)}$ and 
${\rm e}^{-{\rm i}(-K+q)}$ in the equation for $\varepsilon^*(x,t)$ are respectively 
\begin{eqnarray}
-(1+ {\overline q} -\textstyle{\frac{1}{4}}{\overline \omega}^2 )n_+  + \alpha' \varepsilon^{*}_- &=&0,\nonumber\\
-(1 - {\overline q} -\textstyle{\frac{1}{4}} {\overline \omega}^2 )n_-  + \alpha' \varepsilon^*_+ &=&0,\nonumber\\
\left(  V_2- 2{\overline q} + 2{\overline \omega}/{\overline{\rm c}} \right) \varepsilon^{*}_+  - V_2 \varepsilon^{*}_- - {\tilde \alpha} n_- &=&0, \nonumber\\
\left(V_2+ 2{\overline q} + 2{\overline \omega}/{\overline{\rm c}}\right)\varepsilon^{*}_- - {\overline V}_2\varepsilon^{*}_+  - {\tilde \alpha} n_+ &=&0.\label{eq:slowlyvaryA}
\end{eqnarray}
\begin{figure}[h]
\includegraphics[width=.9\linewidth]{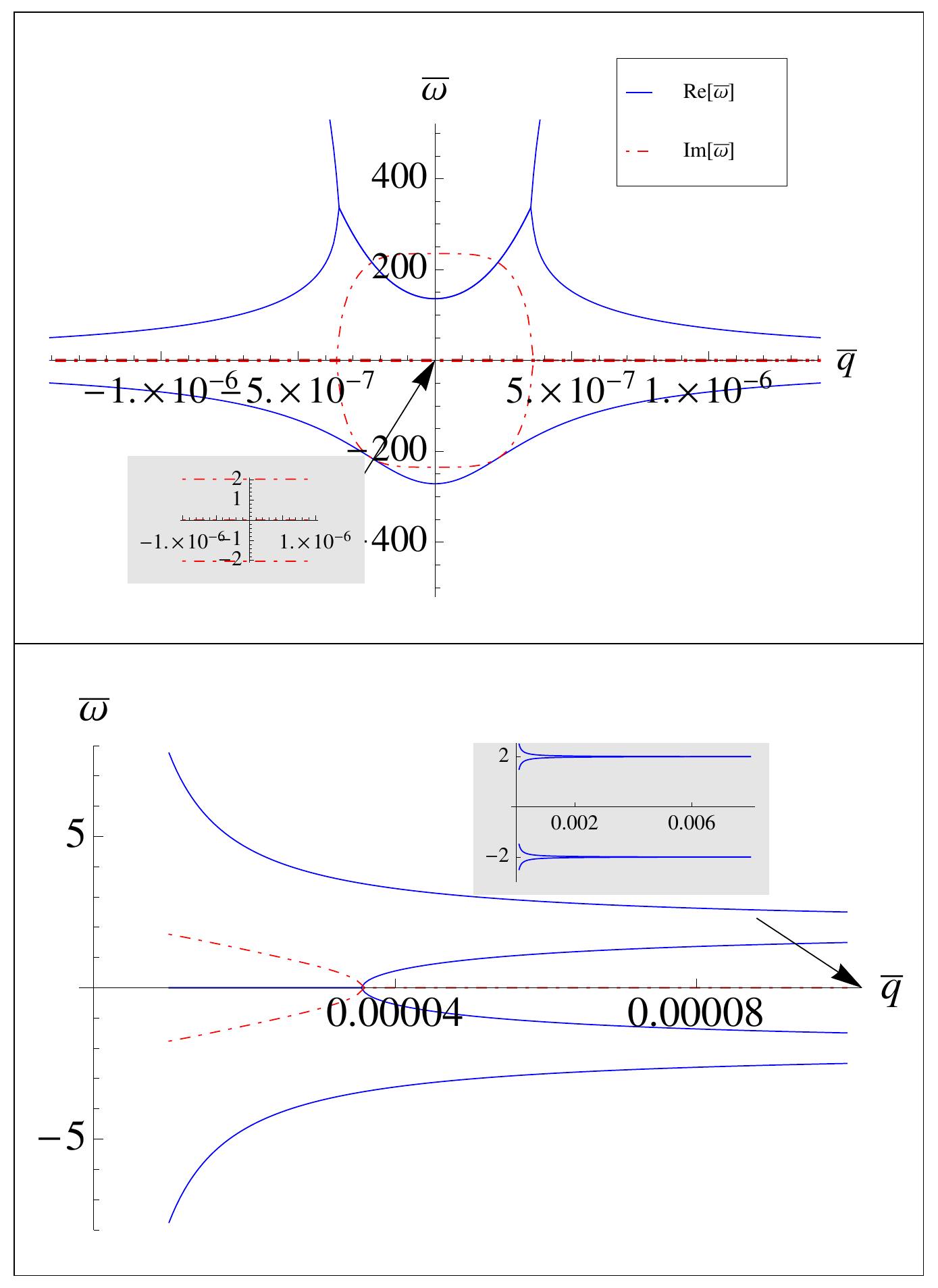}
\caption{Real and imaginary parts of ${\overline \omega}$ for the optical lattice case. The upper inset shows fine detail in $\overline \omega$; the lower insert shows the larger scale structure in $\overline q$. The ``valence band" is off the bottom of the upper figure, curving down. }\label{fig:optl}
\end{figure}
The solutions to these equations are represented in Fig.(\ref{fig:optl}),  we see for sufficiently small ${\overline q}$ the majority of the modes have complex frequencies. To grasp the physical origin of the modes, it is useful to consider ${\overline q}=0$. There one finds there are two stable modes, one of which (not in the Figure because of scale) is the ``conduction band" of the optical lattice, at ${\overline \omega}({\overline q})\simeq -{\overline {\rm c}} (V_2 +  {\overline q}^2/V_2)$, unperturbed from the value obtained with a static distortion, $n_0(x)$, with negligible weight in $n_\pm \sim 1/{\overline {\rm c}}^2$ and $\varepsilon^*_+ = - \varepsilon^*_+$ (as expected). This mode has antinodes at the maxima of $n_0(x)$ and hence is weakly coupled. The other is one member of the three roots ${\overline \omega}({\overline q}=0) = 2 (V_2{\overline {\rm c}})^{1/3}$, with an implicit factor of the three cube roots of unity. They have associated eigenvectors with $\varepsilon^*_+ =  \varepsilon^*_-$, as expected for the ``valence band", with $n_+=n_-$, and small relative weight in the density modes $n_\pm \simeq -\alpha'/(V_2{\overline c})^{2/3} \varepsilon^*_\mp$. Finally there are two roots which are purely imaginary: ${\overline \omega} = \pm 2{\rm i}$, with antisymmetric eigenvectors $\varepsilon^*_+ =- \varepsilon^*_-$, $n_+=-n_-$ and $n_\pm = \frac{1}{2} \alpha' \varepsilon_\mp$, again imbalanced but not so strongly. Note due to the non-Hermiticity of the equations, the modes are not required to be orthogonal. 

As ${\overline q}$ increases, the system becomes stable with the complex roots disappearing in pairs (see Fig.(\ref{fig:optl})).  We find the inner critical wave vector, ${\overline q}^+_{\rm c} \simeq \sqrt{3}(2V_2^2/{\overline c})^{1/3}$,  with ${\overline \omega}_{\rm c} \simeq 2 (2V_2{\overline c})^{1/3}$.    The outer critical wave vector is ${\overline q}^-_{\rm c} \simeq 2V_2$, with ${\overline \omega}_{\rm c} =0$. These values are in very good agreement with the numerical values in the Figure. There are also instabilities\cite{Usf} involving long wavelength phonons, $n(x,t) =  {\tilde n}{\rm e}^{-{\rm i}(qx+\omega t)}$, with no counterpart in the Brillouin instability, which occur on smaller scale in $q$, than those discussed above. We will not consider them further here\cite{Usf}.

We now turn to the observability of these instabilities. The crucial aspect is that they occur at small interval of wave vectors, $q$, around the kinematic point implying a minimum length, $L_{\rm c}$, of fibre (or cavity) for their observation, $L_{\rm c} \gtrsim \pi / q_{\rm c}$. The instability with largest $q_{\rm c}$ will be the most observable. For the optical lattice, this is at $q^-_{\rm c} = K {\overline q}^-_{\rm c} = K \alpha'{\tilde \alpha}/2 $. Using the numerical values quoted below Eqn.(\ref{eq:epsi}), we find $L_{\rm c} \simeq 1$ cm.  Amongst the experimental challenges remaining would be the number of atoms required, $N_{\rm at}$ of perhaps  $10^9$ and the protocol to load the atoms in a subcritical state.  If the laser power is increased $L_{\rm c}$ decreases, but this lies outside the region of validity of our approximations (as $\alpha' \gtrsim 1$). It is natural to conjecture the whole Brillouin Zone is affected as the power increases,  perhaps connecting with the homogeneous heating in previous work\cite{Zol2}.

In summary we have shown that combined instabilities of light and atoms occur in two circumstances: a single beam in conjunction with an initially uniform Bose gas, and a weak optical lattice. These instabilities occur on length scales which are observable. The extension of the theory to the region of stronger optical lattices is presented as a challenge.

\acknowledgements We thank the Aspen Center for Physics, IHP Paris and KITP, UCSB, for a stimulating environment during this work, and K.Bongs, N.R. Cooper, D.M. Gangardt, M.W. Long, M.D. Lukin,  A.J. Schofield and S.F. Yelin for helpful discussions and comments on the manuscript.


\begin{thebibliography}{25}
\expandafter\ifx\csname
natexlab\endcsname\relax\def\natexlab#1{#1}\fi
\expandafter\ifx\csname bibnamefont\endcsname\relax
  \def\bibnamefont#1{#1}\fi
\expandafter\ifx\csname bibfnamefont\endcsname\relax
  \def\bibfnamefont#1{#1}\fi
\expandafter\ifx\csname citenamefont\endcsname\relax
  \def\citenamefont#1{#1}\fi
\expandafter\ifx\csname url\endcsname\relax
  \def\url#1{\texttt{#1}}\fi
\expandafter\ifx\csname urlprefix\endcsname\relax\def\urlprefix{URL
}\fi \providecommand{\bibinfo}[2]{#2}
\providecommand{\eprint}[2][]{\url{#2}}

\bibitem{Ess} F.~Brennecke, T.~Donner, S.~Ritter, T.~Bourdel, M.~K{\" o}hl and T.~Esslinger, Nature {\bf  450}, 268 (2007); K.~Baumann, C.~Guerlin, F.~Brennecke and T.~Esslinger, Nature {\bf  464}, 1301 (2010); K.~Baumann, R.~Mottl, F.~Brennecke, and T.~Esslinger, Phys.\ Rev.\ Lett.\ {\bf 107}, 140402 (2011); H.~Ritsch, P.~Domokos, F.~Brennecke, T.~Esslinger, Rev. Mod. Phys. {\bf 85}  553 (2013).

\bibitem{Dem}D.~E.~Chang, V.~Gritsev, G.~Morig, V.~Vuletic, M.~D.~Lukin and E.~Demler Nature Physics  {\bf 4},884 (2008);  J.~S.~Jin, D.~Rossini, R.~Fazio, M.~Leib and M.~J.~ Hartmann, Phys. Rev. Lett. {\bf 110} 163605 (2013).

\bibitem{Ritsch}W.~Niedenzu, R.~Schulze, A.~Vukics and H.~Ritsch Phys.\ Rev.\ A {\bf 82}, 043605 (2010). 

\bibitem{Zol2} H.~Pichler, A.~J.~Daley and P.~Zoller, Phys. Rev. A {\bf 82}, 063605 (2010).

\bibitem{Ol} M.~A.~Ol'Shanii, Yu.~B.~Ovchinnikov and V.~S.~Letokhov, Opt. Comm. {\bf 98}, 77 (1993).

\bibitem{Zol} S.~Marksteiner, C.~M.~Savage, P.~Zoller and S.~L.~Rolston, Phys. Rev. A {\bf 50}, 2680 (1994).

\bibitem{Cor} M.~J.~Renn, D.~Montgomery, O.~Vdovin, D.~Z.~Anderson, C.~E.~Wieman, and E.~A.~Cornell, Phys. Rev. Lett. {\bf 75} 3253 (1995).


\bibitem{Pr} R.~F.~Cregan, B.~J.~Mangan, J.~C.~Knight, T.~A.~Birks, P.~St.~J.~Russell, P.~J.~Roberts and D.~C.~Allan, Science, {\bf 285}, 1537 (1999).

\bibitem{Rusatt} M.~H.~Frosz, J.~Nold, T.~Weiss, A.~Stefani, F.~Babic, S.~Rammler, and P.~S.~Russell, Opt. Lett. {\bf 38}   2215 (2013);  P.~J.~Roberts, F.~Couny, H.~Sabert, B.~J.~Mangan, D.~P.~Williams, L.~Farr, M.~W.~Mason, A.~Tomlinson, T.~A.~Birks, J.~C.~Knight, and P.~St.~J.~Russell, Opt. Express {\bf 13}, 236 (2005). 

\bibitem{Pr2} F.~Benabid, J.~C.~Knight and P.~St.~J.~Russell, Optics Letters, {\bf 10}, 1195 (2002).

\bibitem{Pr3} T.~.J.~Euser, O.~A.~Schmidt, S.~Unterkofler, and P.~St.~J.~Russell, Optics in the Life Sciences, OSA Technical Digest (online) (Optical Society of America, 2013), paper TT3D.1; T.~.J.~Euser, O.~A.~Schmidt, M.~K.~Gabros, S.~Unterkofler, and P.~St.~J.~Russell, 3rd International Conference on Photonics (2012), 316-317 (2012); O.~A.~Schmidt, M.~K.~Gabros, T.~.J.~Euser and P.~St.~J.~Russell, Optics Letters, {\bf 37}, 91 (2012).

\bibitem{Tak} T.~Takekoshi and R.~J.~Knize, Phys. Rev. Lett. {\bf 98}, 210404 (2007).

\bibitem{Gae}P.~Londero, V.~Venkataraman, A.~R.~Bhagwat, A.~D.~Slepkov, and A.~L.~Gaeta, Phys. Rev. Lett. {\bf 103} 043602 (2009);  V.~Venkataraman, K.~Saha, A.~L.~Gaeta, Nature Photonics  {\bf 7}   138 (2013).

\bibitem{Ket} C.~A.~Christensen, S.~Will, M.~Saba, G-B.~Jo, Y-I.~ Shin, W.~Ketterle, and D.~Pritchard, Phys. Rev.A {\bf 78}, 033429 (2008).

\bibitem{Bac}  M.~Bajcsy, S.~Hofferberth, V.~Balic, T.~Peyronel, M.~Hafezi,
A.~S.~Zibrov, V.~Vuletic, and M.~D.~Lukin, Phys. Rev. Lett. {\bf 102}, 203902 (2009); M.~Bajcsy, S.~Hofferberth, T.~Peyronel, V.~Balic, Q.~Liang, A.~S.~Zibrov, V.~Vuletic, and M.~D.~Lukin, Phys. Rev. A  {\bf 83}, 063830 (2011).

\bibitem{Kai} S.~Vorrath, S.~A.~M\"oller, P.~Windpassinger, K.~Bongs and
K.~Sengstock, New J. Phys., {\bf 12}, 123015 (2010).

\bibitem{Duffing} J.~Kovacic and M.~Brennan, {\em The Duffing Equation}, (Wiley, Oxford, 2011).

\bibitem{NISTFou} F.~W.~J. Olver et al, {\em NIST Handbook of Mathematical Functions}, (Cambridge University Press, 2010), Ch. 22 Eqn. (22.11.2). 
\bibitem{Bill1}I.~H.~Deutsch, R.~J.~C.~Spreeuw, S.~L.~Rolston and W.~D.~Phillips, Phys.\ Rev.\ A, {\bf 52}, 1394 (1995).
\bibitem{Bill2}G.~Birkl,  M.~Gatzke, I.~H.~Deutsch, S.~L.~Rolston and W.~D.~Phillips, Phys.\ Rev.\ Lett. {\bf 75} 2823 (1995).
\bibitem{Usf} N.~K.~Wilkin and J.~M.~F.~Gunn, to be published.





\bibitem{Simm} J.~G.~Simmonds and J.~E.~Mann, {\em A First Look at Perturbation Theory}, (Krieger, Malabar FL, 1986), chapter 2.

\end{thebibliography}
\end{document}